\documentclass[aps,prd,showpacs,superscriptaddress]{revtex4}
\usepackage{amsmath}
\usepackage{graphicx}
\usepackage{bm}
\usepackage{mathrsfs}

\newcommand{\be}{\begin{equation}}
\newcommand{\ee}{\end{equation}}
\newcommand{\ber}{\begin{eqnarray}}
\newcommand{\eer}{\end{eqnarray}}

\def\case#1/#2{\textstyle\frac{#1}{#2} }
\newcommand{\bra}[1]{\left(#1\right)}

\newcommand{\reff}[1]{(\ref{#1})}

\begin{document}

\title{Nonlinear Interactions Between Gravitational Radiation and Modified Alfv\'{e}n Modes in
Astrophysical Dusty Plasmas}
\author{Mats Forsberg}
\affiliation{Department of Physics, Ume{\aa} University,
SE--901 87 Ume{\aa}, Sweden}

\author{Gert Brodin}
\affiliation{Department of Physics, Ume{\aa} University,
SE--901 87 Ume{\aa}, Sweden}
\affiliation{Centre for Fundamental Physics,
 Rutherford Appleton Laboratory,
 Chilton Didcot, Oxfordshire, OX11 0QX, UK} 

\author{Mattias Marklund}
\affiliation{Department of Physics, Ume{\aa} University,
SE--901 87 Ume{\aa}, Sweden}
\affiliation{Centre for Fundamental Physics,
 Rutherford Appleton Laboratory,
 Chilton Didcot, Oxfordshire, OX11 0QX, UK}

\author{Padma K.\ Shukla}
\affiliation{Institut f\"ur Theoretische Physik IV and Centre for Plasma Science and Astrophysics, 
Fakult\"at f\"ur Physik und Astronomie, Ruhr-Universit\"at Bochum, D--44780 Bochum, Germany}

\author{Joachim Moortgat}
\affiliation{Department of Physics \& Astronomy, University of Rochester, Bausch \& Lomb Hall, PO Box 270171, 600 Wilson Boulevard, Rochester, NY 14627-0171, USA.}

\date{\today}

\begin{abstract}
We present an investigation of nonlinear interactions between Gravitational Radiation 
and modified Alfv\'{e}n modes in astrophysical dusty plasmas. Assuming that stationary 
charged dust grains form neutralizing background in an electron-ion-dust plasma,
we obtain the three wave coupling coefficients, and calculate the growth rates for
parametrically coupled gravitational radiation and  modified Alfv\'{e}n-Rao modes. 
The threshold value of the gravitational wave amplitude associated with convective
stabilization is particularly small if the gravitational frequency is close
to twice the modified Alfv\'en wave-frequency. The implication of our results to 
astrophysical dusty plasmas is discussed.
\end{abstract}


\maketitle

\section{Introduction}

There exists several mechanisms for conversion between gravitational waves
(GWs) and electromagnetic waves \cite
{grishchuk,Brodin-Marklund1999,Papadouplous2001,Moortgat2003,Moortgat2004,Servin2003,servinbrodin,
Mendonca2002,Balakin2003,Li2002,BrodinMarklund2003,Picasso2003,Papadoupolus2002,MDB2000,
Hogan2002,bms,Vlahos2004,Mosquera2002}
in plasmas. One of the most basic processes occurs when GWs propagate across
an external magnetic field, which gives rise to a linear coupling to the
electromagnetic field \cite{grishchuk}, leading to the excitation of
magnetohydrodynamic (MHD) waves in a plasma \cite
{Papadouplous2001,Moortgat2003,Moortgat2004}. In order to excite
perturbations with frequencies different from that of the GW, naturally
nonlinear couplings must be considered. There exist numerous examples of
such mechanisms in plasmas, giving rise to, e.g. three-wave couplings
between GWs and electromagnetic waves. Wave coupling mechanisms involving GWs
are studied for several different reasons. In some cases, the emphasis is on
the basic theory \cite{Servin2003,servinbrodin,Mendonca2002,Balakin2003}. In
other works, the focus is on GW detectors \cite
{Li2002,BrodinMarklund2003,Picasso2003}, on cosmology \cite
{Papadoupolus2002,MDB2000,Hogan2002}, or on astrophysical applications such
as binary mergers \cite{bms}, gamma ray bursts \cite{Vlahos2004}, pulsars 
\cite{Mosquera2002} or supernovas \cite{Brodin2005}.

In the present paper, we will consider gravitational wave propagation in
plasmas containing charged dust particles \cite{shukla-mamun}. The latter
are prominent components in many astrophysical systems, and may
contribute significantly to the dynamical properties of such systems \cite
{Mendis-Rosenberg,Horanyi-etal,Okamoto-etal}. It has also been claimed that
supernovae can be significant sources of dust particles \cite{Dunne-etal},
although this claim is debated \cite{Krause-etal}. Previous work involving
dusty plasma-gravitational wave interactions \cite{Brodin2005} have
considered general relativistic versions of the dust MHD equations \cite
{Birk96,shukla-mamun}. However, in cases where the dynamics is not dominated by 
the charged dust particles, other approximations are more useful \cite{Rao95}. In order 
to describe the modified Alfv\'{e}n mode (MAM) (or the  Alfv\'en-Rao mode \cite{Rao95}), 
that can propagate in a magnetized dusty plasma, we will apply the infinite mass approximation 
for immobile charged dust grains. Thus, the only force felt by the charged dust particles 
will be the gravitational force, which is an appropriate approximation for a broad range 
of Alfv\'en wave frequencies. Using the standard mode coupling theory \cite{Weiland-Wilhelmsson}, 
we then obtain  the coupling coefficient describing the nonlinear interaction between 
two MAMs and one GW. Using these results, we consider the parametric excitation of the Rao mode 
in the vicinity of binary mergers. Provided that the gravitational wave frequency is
close to twice the Rao cut-off frequency \cite{Rao95}, the threshold value for the 
GW amplitude for the parametric interaction is much reduced compared to the ideal MHD
theory \cite{Servin2000}, assuming that the limiting amplitude is determined
by non-dissipative stabilization.

\section{Basic equations}

The metric describing a linearized gravitational wave of arbitrary
polarization propagating in the z direction on a flat background can be
written as 
\begin{equation}
		ds^{2}=-dt^{2}+(1+h_{+})dx^{2}+(1-h_{+})dy^{2}+2h_{\times }dxdy+dz^{2}.
\label{metric}
\end{equation}
A convenient choice is to introduce an orthonormal tetrad $e_{i}$, where $%
e_{0}=\partial _{t}$ and the spatial part is written as $\nabla
=(e_{1},e_{2},e_{3})=\nabla _{0}+\nabla _{g}$, where $\nabla $ and $\nabla
_{g}$ are given by 
\begin{eqnarray}
\nabla _{0} &=&(\partial _{x},\partial _{y},\partial _{z}),  \label{nabla_0}
\\
\nabla _{g} &=&-\frac{1}{2}(h_{+}\partial _{x}+h_{\times }\partial
_{y},h_{\times }\partial _{x}-h_{+}\partial _{y},0),  \label{nabla_g}
\end{eqnarray}
and $\mathbf e_1 = (e_1,0,0)$, $\mathbf e_2 = (0,e_2,0)$, $\mathbf e_3 = (0,0,e_3)$.

Below we will derive the evolution equations for the matter and fields in
the low-frequency MHD limit, in the presence of a gravitational wave. We
consider a cold plasma composed  of the electrons, ions and micron-sized
charged dust particles. The momentum equation for the ions, when neglecting pressure and
considering the electrons as massless fluid, reduces to 
\begin{equation}
\left( \partial _{t}+\mathbf{v}_{i}\cdot \nabla \right) \mathbf{v}_{i}=\frac{%
q_{i}}{m_{i}c}(\mathbf{v}_{i}-\mathbf{v}_{e})\times \mathbf{B}+\mathbf{g}%
_{i},  \label{p_i_2}
\end{equation}
where 
\begin{equation*}
\mathbf{g}_{i}=-\frac{1}{2}\left( 1-\frac{v_{iz}}{c}\right) \left[ (v_{ix}%
\dot{h}_{+}+v_{iy}\dot{h}_{\times })\mathbf{e}_{1}+(v_{ix}\dot{h}_{\times
}-v_{iy}\dot{h}_{+})\mathbf{e}_{2}\right] -\frac{1}{2c}\left[
(v_{ix}^{2}-v_{iy}^{2})\dot{h}_{+}+2v_{ix}v_{iy}\dot{h}_{\times }\right] 
\mathbf{e}_{3},
\end{equation*}
is the gravitational acceleration of the ions \cite{tetrad-note}. For low phase speed (in comparison
with the speed of light) the displacement current are neglected, and Maxwell's
equations are written as
\begin{eqnarray}
q_{i}n_{i}\mathbf{v}_{i}-en_{e}\mathbf{v}_{e} &=&\frac{c}{4\pi }\nabla
\times \mathbf{B}-\mathbf{j}_{E},  \label{ampere} \\
\partial _{t}\mathbf{B} &=&\nabla \times \left( \mathbf{v}_{e}\times \mathbf{%
B}\right) -\mathbf{j}_{B},  \label{faraday}
\end{eqnarray}
where 
\begin{eqnarray}
\mathbf{j}_{E} &=&-\frac{1}{2}\left[ (E_{x}-B_{y})\dot{h}_{+}+(E_{y}+B_{x})%
\dot{h}_{\times }\right] \mathbf{e}_{1}+\frac{1}{2}\left[ (E_{y}+B_{x})\dot{h%
}_{+}+(E_{x}-B_{y})\dot{h}_{\times }\right] \mathbf{e}_{2},  \notag \\
\mathbf{j}_{B} &=&-\frac{1}{2}\left[ (E_{y}+B_{x})\dot{h}_{+}-(E_{x}-B_{y})%
\dot{h}_{\times }\right] \mathbf{e}_{1}-\frac{1}{2}\left[ (E_{x}-B_{y})\dot{h%
}_{+}+(E_{y}+B_{x})\dot{h}_{\times }\right] \mathbf{e}_{2},  \notag
\end{eqnarray}
are effective currents due to the GWs, see e.g. Ref. \cite{Servin2003}. 

Combining Eqs. (\ref{p_i_2}) and (\ref{ampere}) and assuming the plasma to
be quasineutral with negatively charged dust, that is $q_{i}n_{i}-en_{e}-q_{d}n_{d}=0$ where $q_d > 0$, we obtain
\begin{equation}
\left( \partial _{t}+\mathbf{v}_{i}\cdot \nabla \right) \mathbf{v}_{i}=\frac{%
q_{i}}{en_{e}m_{i}c}\left( -n_{d}q_{d}\mathbf{v}_{i}+\frac{c}{4\pi }\nabla
\times \mathbf{B}-\mathbf{j}_{E}\right) \times \mathbf{B}+\mathbf{g}_{i}.
\label{v_i}
\end{equation}
By using Eq. (\ref{ampere}) we can eliminate $\mathbf{v}_{e}$ in (\ref{faraday})
to obtain
\begin{equation}
\partial _{t}\mathbf{B}=\nabla \times \left\{ \left( \frac{q_{i}n_{i}}{en_{e}%
}\mathbf{v}_{i}-\frac{c}{4\pi en_{e}}\nabla \times \mathbf{B}+\frac{\mathbf{j%
}_{E}}{en_{e}}\right) \times \mathbf{B}\right\} - \mathbf{j}_{B}.  \label{B}
\end{equation}
Finally, the ion continuity equation is
\begin{equation}
\partial _{t}n_{i}+\nabla \cdot (n_{i}\mathbf{v}_{i})=0.  \label{cont}
\end{equation}
For later use it will be convenient to collect the gravitational terms on
the RHS of the equations. Furthermore, we let the ions to be of charge $
q_{i}=e$. Equations (\ref{v_i}), (\ref{B}) and (\ref{cont}) are then written
as
\begin{equation}
m_{i}n_{e}c\left( \partial _{t}+\mathbf{v}_{i}\cdot \nabla _{0}\right) 
\mathbf{v}_{i}+n_{d}q_{d}\mathbf{v}_{i}\times \mathbf{B}-\frac{c}{4\pi }%
(\nabla _{0}\times \mathbf{B})\times \mathbf{B}=\frac{c}{4\pi }(\nabla
_{g}\times \mathbf{B})\times \mathbf{B}+n_{e}cm_{i}\mathbf{g}_{i}-\mathbf{j}%
_{E}\times \mathbf{B},  \label{v_i_mod}
\end{equation}
\begin{equation}
\partial _{t}\mathbf{B}-\nabla _{0}\times \left[ \frac{n_{i}\mathbf{v}%
_{i}\times \mathbf{B}}{n_{e}}-\frac{c(\nabla _{0}\times \mathbf{B})\times 
\mathbf{B}}{4\pi en_{e}}\right] =\nabla _{g}\times \left[ \frac{n_{i}\mathbf{%
v}_{i}\times \mathbf{B}}{n_{e}}-\frac{c(\nabla _{0}\times \mathbf{B})\times 
\mathbf{B}}{4\pi en_{e}}\right] +\nabla \times \frac{\mathbf{j}_{E}\times 
\mathbf{B}}{en_{e}}-\mathbf{j}_{B},
\end{equation}
and
\begin{equation}
\partial _{t}n_{i}+\nabla _{0}\cdot (n_{i}\mathbf{v}_{i})=-\nabla _{g}\cdot
(n_{i}\mathbf{v}_{i}),  \label{cont_mod}
\end{equation}
where the dust charge is defined by $q_{d}=Z_{d}e$. We note that in the absence of the GW-source terms, eqs. \reff{v_i_mod}-\reff{cont_mod} are MHD equations modified by the presence of infinitely heavy dust particles. Due to the dust charges in the quasi-neutrality condition, the relative contributions to the currents are modified as compared to ideal MHD. As a consequence a characteristic frequency 
$\Omega _{R}$ called the Rao cut-off frequency \cite{Rao95} enters in the linear wave modes. 
Wave modes propagating almost perpendicular to the external magnetic will be denoted as Alfv\'en-Rao modes. In what follows we will be particularly interested in these modes, as they give the simplest way to intoduce a characteristic time scale ($\Omega _{R}^{-1}$), which is longer than the ion Larmor period, into the MHD equations. Furhtermore, as we will demonstrate, these modes are particularly easy to excite when the condition $\omega _{g} = 2\Omega _{R}$ is met, where $\omega _{g}$ is the gravitational wave frequency.

\section{Derivations of the coupled mode equations}

In general a monochromatic wave of sufficient amplitude is subject to a number of instabilities, which can transfer the wave energy into other modes. If energy-momentum conservation allows for resonant three-wave interaction, typically this mechanism give raise to the most rapid instability \cite{Weiland-Wilhelmsson}. Thus, in order to study this process, a system of three weakly interacting waves will be considered. Two dust
MHD waves with frequencies and wavenumbers $(\omega _{1,2},\mathbf{k}%
_{1,2}) $, and a GW with arbitrary polarization propagating parallel to the
background magnetic field with the frequency and the wavenumber $(\omega _{g},%
\mathbf{k}_{g})$. 
We note here that a gravitational wave propagating in a finite angle to the magnetic field produces a linear coupling to the electromagnetic field \cite{grishchuk}. In the next step of the calculations, such linearly induced fields will complicate the description of the nonlinear mode coupling to a large extent. Thus our motive to let $\mathbf k_g$ be parallel to the external magnetic field is to avoid a linear coupling between the GW and the MHD modes, and be able to focus on the nonlinear phenomena.
As a prerequisite to obtain the nonlinear coupling, we first study the
linear modes of the system (\ref{v_i_mod})-(\ref{cont_mod}), omitting the
gravitational contributions. Considering the plasma waves to be plane wave
perturbations on background quantities such that $\mathbf{B}=B_{0}\hat{%
\mathbf{z}}+\mathbf{B}^{\prime }$, $n_{i}=n_{0}+n^{\prime }$ and $\mathbf{v}%
_{i}=\mathbf{v}^{\prime }$, where $B_{0}$ and $n_{0}$ are constant and the
primed quantities denote the perturbations, allows us to write the linear
part of (\ref{v_i_mod})-(\ref{cont_mod}) as 
\begin{eqnarray}
\mathbf{v}^{\prime } &=&-i\frac{\Omega _{R}}{\omega }\mathbf{v}^{\prime
}\times \hat{\mathbf{z}}-\frac{C_{A}}{\sqrt{4\pi n_{0}m_{i}}}\frac{[k_{z}%
\mathbf{B}^{\prime }-B_{z}^{\prime }\mathbf{k}]}{\omega },  \label{primed-a}
\\
\mathbf{B}^{\prime } &=&\sqrt{4\pi n_{0}m_{i}}C_{A}\frac{[(\mathbf{k}\cdot 
\mathbf{v}^{\prime })\hat{\mathbf{z}}-k_{z}\mathbf{v}^{\prime }]}{\omega }+%
\frac{im_{i}cC_{A}}{e\sqrt{4\pi n_{0}m_{i}}}\frac{k_{z}\mathbf{k}\times 
\mathbf{B}^{\prime }}{\omega },  \label{primed-b} \\
n^{\prime } &=&\frac{n_{0}}{\omega }\mathbf{k}\cdot \mathbf{v}^{\prime },
\label{primed-c}
\end{eqnarray}
where $\Omega _{R}=n_{d}q_{d}B_{0}/m_{i}(n_{0}-Z_{d}n_{d})c$ is the Rao cut-off frequency and $C_{A}=n_{0}B_{0}/(n_{0}-Z_{d}n_{d})\sqrt{%
4\pi n_{0}m_{i}}$ is the Alfv\'{e}n speed.
The frequency matching is
\begin{equation}
\omega _{g}=\omega _{1}+\omega _{2},  \label{frequency-match}
\end{equation}
and since the gravitational dispersion relation reads $\omega
_{g}=ck_{g}$ and $C_{A}\ll c$, the wavenumber matching can be approximated by 
\begin{equation}
\mathbf{k}_{g}=\mathbf{k}_{1}+\mathbf{k}_{2}\Rightarrow \mathbf{k}%
_{1}\approx -\mathbf{k}_{2}.  \label{wave-vector-match}
\end{equation}
Thus we may consider the excitation of MHD wave-modes with wave-vectors that are almost perpendicular to the GW (or the external magnetic field). In particular we choose a coordinate system such that $k_{y}=0$ and let $|k_{1,2z}|\ll|k_{1,2x}|$, 
which allows the linear eigenmodes of the system to be represented by the following eigenvector 
\begin{equation}
\left( 
\begin{array}{c}
v_{x}^{\prime } \\ 
v_{y}^{\prime } \\ 
B_{z}^{\prime } \\ 
n^{\prime }
\end{array}
\right) =v_{x}^{\prime }\left( 
\begin{array}{c}
1 \\ 
i\frac{\Omega _{R}}{\omega } \\ 
\frac{C_{A}\sqrt{4\pi n_{0}m_{i}}}{\omega }k_{x} \\ 
\frac{n_{0}}{\omega }k_{x}
\end{array}
\right) .  \label{linear_eigenvector}
\end{equation}
The dispersion relation (DR) is now readily obtained and can be expressed as
\begin{equation}
\omega ^{2}=\Omega _{R}^{2}+C_{A}^{2}k^{2},  \label{DR}
\end{equation}
which is the Alfv\'{e}n-Rao mode \cite{Rao95}, that reduces to the
compressional Alfv\'{e}n 
(or fast magnetosonic) wave in the limit of zero dust-density. We note that
the dispersion relation (\ref{DR}), together with (\ref{frequency-match}) and
(\ref{wave-vector-match}), implies $\omega _{g}=2\omega _{1}=2\omega _{2}$.

Next, assuming the dust MHD waves and the GWs to be plane waves with weakly
varying amplitudes, we write $h_{+,\times }=\tilde{h}_{+,\times }(t)e^{i(%
\mathbf{k}_{g}\cdot \mathbf{z}-\omega _{g}t)}+\mathrm{c.c}.$ and $\psi
^{\prime }=\widetilde{\psi }(t)e^{i(\mathbf{k}_{1,2}\cdot \mathbf{r}-\omega
_{1,2}t)}+\mathrm{c.c}$, where $\mathrm{c.c}$ stands for complex conjugate,
and $\psi ^{\prime }$ represents any component of $\mathbf{B}^{\prime },%
\mathbf{v}^{\prime }$ and $n^{\prime }$. Making use of the linear
eigenvector (\ref{linear_eigenvector}) as approximations in the nonlinear
terms in the system (\ref{v_i_mod})-(\ref{cont_mod}), and keeping only the
resonant part to second order in the amplitudes, we obtain the coupled mode
equations \cite{Weiland-Wilhelmsson} 
\begin{equation}
\partial _{t}\tilde{v}_{1,2x}=i\frac{\omega _{g}}{4}\left( 1+\frac{4\Omega
_{R}^{2}}{\omega _{g}^{2}}\right) \tilde{v}_{2,1x}^{\ast }\tilde{h}%
_{+}+2\Omega _{R}\tilde{v}_{2,1x}^{\ast }\tilde{h}_{\times }.
\label{v_1_coupling}
\end{equation}
For the GWs we obtain, by using Einsteins equations linearized in $%
h_{+},h_{\times }$, and keeping only the resonant part of the
energy-momentum tensor 
\begin{equation}
\partial _{t}\tilde{h}_{+}=i\frac{\kappa }{\omega _{g}}m_{i}n_{0}\left( 1+%
\frac{4\Omega _{R}^{2}}{\omega _{g}^{2}}\right) \tilde{v}_{1x}\tilde{v}_{2x},
\label{+-polarization}
\end{equation}
and 
\begin{equation}
\partial _{t}\tilde{h}_{\times }=-4\kappa m_{i}n_{0}\frac{\Omega _{R}}{%
\omega _{g}^{2}}\tilde{v}_{1x}\tilde{v}_{2x},  \label{x-polarization}
\end{equation}
for the $+$ and $\times $-polarization respectively.

Noting that the gravitational wave energy density can be written as 
\begin{equation}
E_{g}\equiv W_{g}\left( |h_{+}|^{2}+|h_{\times }|^{2}\right) =\frac{\omega
_{g}^{2}}{2\kappa }\left( |h_{+}|^{2}+|h_{\times }|^{2}\right),
\label{GW-energy}
\end{equation}
and the Alfv\'{e}n-Rao wave energy density as 
\begin{equation}
E_{1,2}\equiv W_{1,2}|v_{1,2x}|^{2}=\frac{1}{2}m_{i}n_{0}\left(
|v_{1,2x}|^{2}+|v_{1,2y}|^{2}\right) +\frac{|B_{1,2z}|^{2}}{8\pi }%
=m_{i}n_{0}|v_{1,2x}|^{2},  \label{wave-energy}
\end{equation}
we can deduce three independent conservation laws from the coupled-mode
equations (\ref{v_1_coupling})-(\ref{x-polarization}). For the total wave
energy, i.e. 
\begin{equation}
\frac{d(E_{g}+E_{1}+E_{2})}{dt}=0,  \label{E-cons}
\end{equation}
the difference in Alfv\'{e}n-Rao wave quanta, i.e. 
\begin{equation}
\frac{d(N_{1}-N_{2})}{dt}=0,  \label{Q1-cons}
\end{equation}
and the sum of wave quanta, i.e. 
\begin{equation}
\frac{d(2N_{g}+N_{1}+N_{2})=0}{dt},  \label{Q2-cons}
\end{equation}
where we have introduced the number density of gravitational wave quanta $%
N_{g}=E_{g}/\hbar \omega _{g}$ and of the Alfv\'{e}n-Rao wave quanta $%
N_{1,2}=E_{1,2}/\hbar \omega _{1,2}$. The existence of these conservation
laws are equivalent to the Manley-Rowe relations \cite{Weiland-Wilhelmsson}.

For simplicity, we have made the derivation of the coupled mode equations
considering only time-dependent amplitudes. However, we note that a
generalization to allow for weakly space-dependent amplitudes can be made by
the simple substitution $\partial _{t}\rightarrow \partial _{t}+\mathbf{v}%
_{g}\cdot \nabla $ for each mode \cite{Weiland-Wilhelmsson}, i.e. for a
general slowly varying amplitude the coupled mode equations reads 
\begin{equation}
\left( \partial _{t}+v_{g1,2}\partial _{x}\right) \tilde{v}_{1,2x}=i\frac{%
\omega _{g}}{4}\left( 1+\frac{4\Omega _{R}^{2}}{\omega _{g}^{2}}\right) 
\tilde{v}_{2,1x}^{\ast }\tilde{h}_{+}+2\Omega _{R}\tilde{v}_{2,1x}^{\ast }%
\tilde{h}_{\times },  \label{space-1}
\end{equation}
\begin{equation}
\left( \partial _{t}+c\partial _{z}\right) \tilde{h}_{+}=i\frac{\kappa }{%
\omega _{g}}m_{i}n_{0}\left( 1+\frac{4\Omega _{R}^{2}}{\omega _{g}^{2}}%
\right) \tilde{v}_{1x}\tilde{v}_{2x},  \label{space-2}
\end{equation}
and 
\begin{equation}
\left( \partial _{t}+c\partial _{z}\right) \tilde{h}_{\times }=-4\kappa
m_{i}n_{0}\frac{\Omega _{R}}{\omega _{g}^{2}}\tilde{v}_{1x}\tilde{v}_{2x},
\label{space-3}
\end{equation}
where $v_{g1,2}=k_{1,2x}C_{A}^{2}/\omega _{1,2},$ where $k_{1,2x}$ is the
x-component of the wavevector for mode 1 and 2 respectively \cite{Sign-note}.

\section{Summary and discussion}

We have considered the nonlinear interaction between the modified Alfv\'{e}n 
(or the Alfv\'en-Rao) mode and gravitational waves in a magnetized dusty plasma. 
In order to describe this process, dust MHD equations incorporating the effects of
 the gravitational waves have been derived. In particular, we 
have focused on the case where a gravitational wave of arbitrary polarization propagates
parallel to the magnetic field. We have then calculated the three-wave
coupling coefficients for MHD waves propagating almost perpendicular to the
magnetic field,
in which case the latter wave modes obey the Alfv\'{e}n-Rao
dispersion relation (\ref{DR}). From the coupled mode equations we note that the GW can parametrically 
excite Alfv\'{e}n-Rao modes, which grow as $\exp(\gamma t)$, where $\gamma$ depend on $\Omega_R$, $\omega _{g}$, $\tilde{h}_{+,\times}$ etc. Furthermore, it can be seen that $\gamma$ is roughly independent of the GW polarization,
and of the order $\gamma \sim |\tilde{h}_{+,\times }|\omega _{g}$.
This estimate is the same as for the decay into high-frequency waves, see Ref 
\cite{Brodin-Marklund1999}, or MHD waves, see e.g. Refs. \cite
{Papadouplous2001,Servin2000}.
The highest growth rate may be reached for approximately monochromatic
gravitational waves, which could be produced by compact
binaries close to merging, see e.g. \cite{Servin2000}, or during the black-hole ringdown \cite{Chris2004}.

However, we note that
the high gravitational amplitudes only exist during a limited time, and that
the finite group velocity of the decay products has a stabilizing effect on
the parametric process. To study this effect we consider the decay of a
homogeneous intense gravitational wave which for definiteness has the
polarization $\tilde{h}=\tilde{h}_{+}$ . The amplitudes of the
Alfv\'{e}n-Rao modes are assumed to have the form $\tilde{v}_{1x}=\hat{v}%
_{1x}\exp (iKx-i\Omega t)$ and $\tilde{v}^*_{2x}=\hat{v}^*_{2x}\exp(iKx-i\Omega t)$. Inserting this ansatz latter into Eq. (\ref{space-1}), we immediately obtain the nonlinear dispersion relation 
\begin{equation}
		(\Omega -v_{g1}K)(\Omega -v_{g2}K)=-\frac{\omega _{g}^{2}}{16}
		\bra{ 1 + \frac{4 \Omega^2_R}{\omega_g^2} }^2 \left| \tilde{h}_{+}\right| ^{2}.	\label{Growth-rate-DR}
\end{equation}
Next, introducing $v_{g}=\left| v_{g1}\right| =\left| v_{g2}\right| $, we
deduce from (\ref{Growth-rate-DR}) that the growth rate of the
Alfv\'{e}n-Rao modes is 
\begin{equation}
		\gamma =\frac{1}{4}\sqrt{\omega _{g}^{2}\bra{ 1 + \frac{4\Omega^2_R}{\omega_g^2} }^2
		\left| \tilde{h}_{+}\right| ^{2}-16{K^{2}v}_{g}^{2}},  \label{Growth-rate-stabilizing}
\end{equation}
provided that the second term under the root sign is smaller than the first one.
While Eq. (\ref{Growth-rate-stabilizing}) formally shows that sufficiently
long wavelength amplitude perturbations always are unstable, in reality
there is a minimum wavenumber possible. This wavenumber $K_{\min }$ is set
by the shortest non-oscillatory scale of the problem, which may be either
the inverse inhomogeneity scalelength $\nabla n_{0}/n_{0}$, $R_{\mathrm{curv%
}}^{-1}$ or ($ct_{p})^{-1}$, where $R_{\mathrm{curv}}$ is the background
curvature and $t_{p}$ is the pulse duration time. Note that the pulse
duration here refers to the time during which the pulse has a sufficiently
constant frequency, such as not to break the frequency matching condition.
The threshold value on the gravitational wave amplitude for the parametric
excitation now becomes 
\begin{equation}
\tilde{h}_{+\mathrm{tre}}=\frac{8{K}_{\min }v_{g}}{\omega _{g}}.
\label{treshold}
\end{equation}
Since the group velocity of the Alfv\'{e}n-Rao mode approaches zero when $%
\omega \rightarrow \Omega _{R}$, we note that $\tilde{h}_{+\mathrm{tre}%
}\rightarrow 0$ when $\omega _{g}\rightarrow 2\Omega _{R}$. Introducing the
frequency mismatch $\delta \omega =\omega _{g}-2\Omega _{R}$, and using Eq. (%
\ref{treshold}) we find that the threshold value is 
\begin{equation}
\tilde{h}_{+\mathrm{tre}}=\frac{8\delta \omega }{\omega _{g}}{K}_{\min }C_{A}.
\label{threshold-2}
\end{equation}
Thus, we conclude that provided the gravitational spectrum contains twice the
Rao-frequency, parametric excitation of MHD waves occur more easily in a dusty plasma, that may exist in astrophysical systems, see e.g. Refs. 
\cite{Mendis-Rosenberg,Horanyi-etal,Okamoto-etal,Dunne-etal}, as compared to a plasma without dust-particles.
 As a consequence of the low threshold,
GW-induced Alfv\'{e}n-Rao modes may be excited at a comparatively large
distance from the gravitational wave source.
The aim of this study has been to shed further light on the gravitational-MHD interactions that take place
in the vicinity of gravitational wave sources, such as collapsing compact binaries, quaking neutron stars, black holes during ringdown or supernovas \cite{supernova}.

\acknowledgments
This research was partially supported by the Swedish Research Council and the Swedish graduate school of space technology.


\begin{thebibliography}{99}
\bibitem{tetrad-note} When including the nonlinear GW-coupling, it is in principle important to separate the coordinate components (indices $x,y,z$) from the tetrad components (indices $1,2,3$), since the difference is first order in $h_+,h_\times$. However, for notational convenience we let indices $x,y,z$ denote tetrad components $1,2,3$ in what follows.
\bibitem{Sign-note}  Note that the group velocities have equal magnitudes
but different signs due to Eq. (\ref{wave-vector-match})
\end{thebibliography}
\end{document}